\documentstyle[12pt]{article}
\begin{document}
\begin{center}
\Large{\bf Global Monopoles in Brans-Dicke} \\ {\bf Theory of Gravity}
\end{center}
\begin{center}
\vspace*{1.5cm}
A. Barros \\ 
{\rm Departamento de F\'\i sica, Universidade Federal de Roraima, \\ 
69310-270, Boa Vista, RR - Brazil.}\\
and\\
C. Romero\footnote{\rm Electronic address: cromero@dfjp.ufpb.br} 
\\
\rm Departamento de F\'\i sica, Universidade Federal da Para\'\i ba, \\ 
Caixa Postal 5008, 58059-970, Jo\~ao Pessoa, PB - Brazil.
\end{center}

\vspace*{1.5cm}
\begin{center} {\bf Abstract}
\end{center}
%{\footnotesize \rm 
{The gravitational field of a global monopole in the context of Brans-Dicke theory of gravity is investigated. The space-time and the scalar field generated by the monopole are obtained by solving the field equations in the weak field approximation. A comparison is made with the corresponding results predicted by General Relativity.}

%\pacs{PACS numbers: 1127, 0450, 9880C}
\newpage
\renewcommand{\thesection}{\Roman{section}.} 
$         $

Monopoles resulting from the breaking of global $O(3)$ symmetry lie among those strange and exotic objects like cosmic strings and domain walls \cite{1}, generally referred to as topological defects of space-time, which may have existed due to phase transitions in the early universe. Likewise cosmic strings, the most studied of these structures, the gravitational field of a monopole exhibits some interesting properties, particularly those concerning the appearance of nontrivial space-time topologies.

The solutions corresponding to the metrics generated by strings \cite{2}, domain walls \cite{2} and global monopoles \cite{3} in the context of General Relativity were all first obtained using the weak field approximation.

In a similar approach, the gravitational fields of cosmic strings and domain walls have been obtained regarding Brans-Dicke theory of gravity and more general scalar-tensor theories of gravity \cite{4,5}.

In this paper we consider the global monopole and investigate its gravitational field by working out Brans-Dicke equations using once more the weak field approximation, essentially in the same way as in the previous works mentioned above.
\renewcommand{\thesection}{\Roman{section}.}

Let us consider Brans-Dicke field equations in the form
\begin{eqnarray}
\label{2.1}
R_{\mu \nu} = {8\pi\over \phi}\left[T_{\mu \nu} - {g_{\mu \nu}\over 2}\left({2\omega +2\over 2\omega +3}\right)T\right] + {\omega \over \phi^2}\phi_{,\mu}\phi_{,\nu} + {1\over \phi}\phi_{;\mu;\nu}\hspace{.2cm},
\end{eqnarray}
\begin{eqnarray}
\label{2.2}
\Box \phi = {8\pi T\over 2\omega +3}\hspace{.2cm},
\end{eqnarray}
where $\phi$ is the scalar field, $\omega$ is a dimensionless coupling constant and $T$ denotes the trace of $T^{\mu}_{\nu}$--- the energy-momentum tensor of the matter fields.

The energy-momentum tensor of a static global monopole can be 
appro\-xi\-ma\-ted (outside the core) as \cite{3} 
\begin{eqnarray}
\label{2.3}
T^{\mu}_{\nu} = \hbox{diag}\left({\eta^2\over r^2}, {\eta^2\over r^2}, 0, 0\right),
\end{eqnarray}
where $\eta$ is the energy scale of the symmetry breaking.

Due to spherical symmetry we consider $\phi = \phi (r)$ and the line element
\begin{eqnarray}
\label{2.4}
ds^2 = B(r)dt^2 - A(r)dr^2 - r^2(d\theta^2 + \sin^2\theta d\varphi^2).
\end{eqnarray}

Substituting this into Eq. (\ref{2.1}) and Eq. (\ref{2.2}), and taking in account Eq. (\ref{2.3}) we obtain the following set of equations: 
\begin{eqnarray}
\label{2.5}
{B''\over 2A} - {B'\over 4A}\left({A'\over A}+{B'\over B}\right)+{1\over r}{B'\over A} = {8\pi\over \phi}\left[{\eta^2B\over r^2(2\omega +3)}\right]-{B'\phi'\over 2A\phi}\hspace{.2cm},
\end{eqnarray}
\begin{eqnarray}
\label{2.6}
-{B''\over 2B} + {B'\over 4B}\left({A'\over A}+{B'\over B}\right)+{1\over r}{A'\over A}=&-&{8\pi\over \phi}\left[{\eta^2A\over r^2(2\omega +3)}\right]+{\omega\phi'^2\over \phi^2}\nonumber \\
&+&{1\over \phi}\left[\phi''-{A'\over 2A}\phi'\right]\hspace{.2cm},
\end{eqnarray}
\begin{eqnarray}
\label{2.7}
\phi'' + {1\over 2}\phi'\left[{B'\over B}-{A'\over A}+{4\over r}\right]= -{16\pi\over (2\omega +3)}\left({\eta^2\over r^2}\right)A\hspace{.2cm},
\end{eqnarray}
\begin{eqnarray}
\label{2.8}
1 - {r\over 2A}\left({B'\over B}-{A'\over A}\right)-{1\over A}= {8\pi\over \phi}\left[\eta^2\left({2\omega +2\over 2\omega+3}\right)\right]+{r\phi'\over A\phi}\hspace{.2cm},
\end{eqnarray}
where prime denotes differentiation with respect to $r$.

Now, dividing (\ref{2.5}) and (\ref{2.6}) by $B$ and $A$, respectively, and adding we get
\begin{eqnarray}
\label{2.9}
{\alpha\over r} = {\omega \phi'^2\over \phi^2} +{\phi''\over \phi}-{\phi'\over 2\phi}\alpha \hspace{.2cm},
\end{eqnarray}
where we have put 
\begin{eqnarray}
\label{2.10}
\alpha = {A'\over A}+{B'\over B}.
\end{eqnarray}

Then, equations (\ref{2.7}) and (\ref{2.8}) read
\begin{eqnarray}
\label{2.11}
\phi'' + {\phi'\over 2}\left[\alpha - {2A'\over A} + {4\over r}\right] = -{16\pi\over 2\omega +3}\left({\eta^2\over r^2}\right)A\hspace{.2cm},
\end{eqnarray}
\begin{eqnarray}
\label{2.12}
1-{r\over 2A}\left(\alpha - {2A'\over A}\right)-{1\over A} = {8\pi\over \phi}\left[\eta^2\left({2\omega +2\over 2\omega+3}\right)\right] +{r\over A}{\phi'\over \phi}.
\end{eqnarray}

At this stage, let us consider the weak field approximation and assume that \\
$A(r)=1+f(r),\qquad B(r)=1+g(r)$\qquad and \qquad $\phi(r)=\phi_o+\epsilon(r)$,\\ 
where $\phi_o$ is a constant which may be identified to $G^{-1}$ when $\omega \rightarrow  \infty$  ($G$ being the Newtonian gravitational constant), and the functions $f, g$ and ${\epsilon \over \phi_o}$ should be computed to first order in ${\eta^2\over \phi_o}$, with $|f(r)|,\hspace{.2cm}|g(r)|,\hspace{.2cm}\left|{\epsilon(r)\over \phi_o}\right| \ll 1$.

In this approximation it is easy to see that 
\begin{eqnarray}
{\phi'\over \phi}={\epsilon'\over \phi_o[1+\epsilon/\phi_o]}= {\epsilon'\over \phi_o}\hspace{.2cm}, \qquad{\phi''\over \phi} = {\epsilon''\over \phi_o[1+\epsilon/\phi_o]}= {\epsilon''\over \phi_o}\hspace{.2cm}, \nonumber
\end{eqnarray}
\begin{eqnarray}
{B'\over B} = {g'\over 1+g}=  g', \qquad {A'\over A} = {f'\over 1+f} = f', 
\nonumber 
\end{eqnarray}
and so on.

From equation (\ref{2.9}) it follows that
\begin{eqnarray}
\label{2.13}
{\alpha \over r} = {\epsilon''\over \phi_o}.
\end{eqnarray}
And from (\ref{2.11}) we have
\begin{eqnarray}
\label{2.14}
\epsilon'' + {2\epsilon' \over r} = -{16\pi\over (2\omega +3)}{\eta^2\over r^2} \hspace{.2cm},
\end{eqnarray}
the solution of which is given by
\begin{eqnarray}
\label{2.15}
\epsilon =  -{16\pi\over 2\omega +3}\eta^2 \ln {r\over r_o} - {\kappa\over r}\hspace{.2cm},
\end{eqnarray}
$r_o$ and $\kappa$ being integration constants.

On the other hand, considering Eq. (\ref{2.13}) and Eq. (\ref{2.15}), equation (\ref{2.12}) becomes
\begin{eqnarray}
\label{2.16}
f' + {f\over r} = {16\pi \eta^2\over \phi_o(2\omega +3)r}(\omega + {1\over 2})
\hspace{0.2cm},
\end{eqnarray}
which yields the solution
\begin{eqnarray}
\label{2.17}
f = {8\pi \eta^2 (2\omega +1)\over \phi_o(2\omega+3)}+{l\over r}\hspace{.2cm},
\end{eqnarray}
where $l$ is an arbitrary constant.

Therefore,
\begin{eqnarray}
\label{2.18}
A = 1+ f = 1+ {8\pi \eta^2 (2\omega +1)\over \phi_o(2\omega +3)}+{l\over r}
\end{eqnarray}
and
\begin{eqnarray}
A^{-1} = 1- {8\pi \eta^2 (2\omega +1)\over \phi_o(2\omega +3)}-{l\over r}.
\end{eqnarray}

It is currently known that solutions of Brans-Dicke field equations do not always go over General Relativity solutions when $\omega \rightarrow \infty$ \cite{6}. However, as the term ${\omega \phi_{,\mu}\phi_{,\nu}\over \phi^2}$ in equation (\ref{2.1}) is neglected in the weak field approximation we expect that in the limit $\omega \rightarrow \infty$ our solution reduces to Barriola-Vilenkin space-time, which is given by \cite{3}
\begin{eqnarray}
\label{2.20}
ds^2 = & &\left(1-8\pi G\eta^2 - {2GM\over r}\right)dt^2 - \left(1-8\pi G\eta^2 - {2GM\over r}\right)^{-1}dr^2
  \nonumber \\
& &- r^2(d\theta^2 + \sin^2 \theta d\varphi^2).
\end{eqnarray}

Then, we should have 
\begin{eqnarray}
\lim _{\omega \rightarrow  \infty} l = 2GM,
\end{eqnarray}
where $M$ is the mass of the monopole core. Indeed, if we take $\eta=0$ in a region outside the monopole core, then a simple comparison of the 
$r$-dependent term in (\ref{2.18}) with the corresponding term of Brans-Dicke 
solution for a spherically symmetric matter distribution in the weak field 
approximation \cite{7}, which may be written as 
\begin{eqnarray}
\label{2.22}
ds^2 = & &\left[1-{2M\over r\phi_o}\left(1 + {1\over 2\omega +3}\right)\right]dt^2 - \left[1+{2M\over r\phi_o}\left(1 - {1\over 2\omega +3}\right)\right]dr^2 \nonumber \\
& &
-r^2(d\theta^2 + \sin^2 \theta d\varphi^2)\hspace{.2cm},
\end{eqnarray} 
gives  $ l = {2M\over \phi_o}\left[1 - {1\over 2\omega +3}\right].$

The same argument concerning the scalar field leads us to $\kappa = -{2M\over 2\omega +3}$. Thus, we have 
\begin{eqnarray}
\label{2.23}
A = 1 + {8\pi \eta^2\over \phi_o}\left({2\omega +1\over 2\omega +3}\right) + {2M\over r\phi_o}\left(1 - {1\over 2\omega +3}\right),
\end{eqnarray}
\begin{eqnarray}
\label{2.24}
\phi = \phi_o - {16\pi \eta^2\over 2\omega +3}\ln {r\over r_o} + {2M\over (2\omega +3)r}.
\end{eqnarray}

From Eq. (\ref{2.10}) it is straightforward to verify that
\begin{eqnarray}
\label{2.25}
B = {a\over A}\left[1-{4M\over r\phi_o(2\omega +3)} + {16\pi \eta^2\over \phi_o(2\omega +3)}\ln {r\over r_o}\right],
\end{eqnarray}
where $a$ is an integration constant. For convenience let us rescale the time by putting $a=1 -{16\pi \eta^2\over \phi_o(2\omega +3)}$. Then,
\begin{eqnarray}
\label{2.26}
B = {1\over A}\left[1-{4M\over r\phi_o(2\omega +3)} + {16\pi \eta^2\over \phi_o(2\omega +3)}\ln {r\over r_o}\right]\left[1 - {16\pi \eta^2\over \phi_o(2\omega +3)}\right].
\end{eqnarray}

Taking into account (\ref{2.23}) we obtain
\begin{eqnarray}
\label{2.27}
B = 1 - {8\pi \eta^2\over \phi_o} +  {16\pi \eta^2\over \phi_o(2\omega +3)}\ln {r\over r_o} - {2M\over r\phi_o}\left[1+{1\over 2\omega +3}\right] .
\end{eqnarray}

Following Barriola-Vilenkin's reasoning we drop the mass term in (\ref{2.23}), (\ref{2.24}) and (\ref{2.27}) as it is negligible on the astrophysical scale. Thus, we have finally
\begin{eqnarray}
\label{2.28}
A(r) = 1 + {8\pi \eta^2 (2\omega +1)\over \phi_o(2\omega +3)}\hspace{0.2cm},
\end{eqnarray}
\begin{eqnarray}
\label{2.29}
B(r) = 1 - {8\pi \eta^2 \over \phi_o} + {16\pi \eta^2\over \phi_o(2\omega +3)}\ln {r\over r_o}\hspace{0.2cm},
\end{eqnarray}
\begin{eqnarray}
\label{2.30}
\phi(r) = \phi_o - {16\pi \eta^2\over 2\omega +3}\ln {r\over r_o}.
\end{eqnarray}

It is not difficult to show that the line element defined by the functions $A(r)$ and $B(r)$ above is conformally related to the Barriola-Vilenkin monopole solution. To do so, let us consider the coordinate transformation given by the equations

\begin{eqnarray}
\label{2.31}
B(r) = h(r^*)\left(1-{8\pi \eta^2\over \phi_o}\right),
\end{eqnarray}
\begin{eqnarray}
\label{2.32}
A(r)dr^2 = h(r^*)\left(1+{8\pi \eta^2\over \phi_o}\right)dr^{*2},
\end{eqnarray}
\begin{eqnarray}
\label{2.33}
r = h^{1/2}(r^*)r^*,
\end{eqnarray}
where $h(r^*)$ is to be calculated and $h(r^*) = 1 + q(r^*)$, with $|q(r^*)| \ll 1$.

Differentiating (\ref{2.33}) we obtain 
\begin{eqnarray}
\label{2.34}
dr^2 = (1 + \dot{q}r^* + q)dr^{*2},
\end{eqnarray}
where dot stands for derivative with respect to $r^*$.

Substituting (\ref{2.34}) into (\ref{2.32}) one gets
\begin{eqnarray}
\label{2.35}
q(r^*)={16\pi \eta^2\over \phi_o(2\omega +3)}\ln{r^*\over r_o}\hspace{0.2cm},
\end{eqnarray}
whence
\begin{eqnarray}
\label{2.36}
h(r^*)=1 + {16\pi \eta^2\over \phi_o(2\omega +3)}\ln{r^*\over r_o}.
\end{eqnarray}

In order to verify the consistency of this result with (\ref{2.29}) let us calculate $B(r)$ directly from (\ref{2.31}) and (\ref{2.36}). Keeping only linear terms in ${\eta^2\over \phi_o}$ and using (\ref{2.33}), we have then
\begin{eqnarray}
\label{2.37}
B(r) &=& \left(1+{16\pi \eta^2\over \phi_o(2\omega +3)}\ln{r^*\over r_o}\right)\left(1-{8\pi \eta^2\over \phi_o}\right)\nonumber \\&=& 1-{8\pi \eta^2\over \phi_o}+{16\pi \eta^2\over \phi_o(2\omega +3)}\ln{r\over r_o}.
\end{eqnarray}

Therefore, the line element (\ref{2.4}) which represents the space-time generated by the monopole may be written in terms of the new coordinate $r^*$ as
\begin{eqnarray}
\label{2.38}
ds^2 = & &\left(1+{16\pi \eta^2\over \phi_o(2\omega +3)}\ln{r^*\over r_o}\right)\biggr[\left(1-{8\pi \eta^2\over \phi_o}\right)dt^2 - \left(1+{8\pi \eta^2\over \phi_o}\right) dr^{*2}  \nonumber \\
& &-r^{*2}\left(d\theta^2 + \sin^2\theta d\varphi^2\right)\biggr] .
\end{eqnarray}

Rescaling the time and defining a new radial coordinate $r=\left(1+{4\pi \eta^2\over \phi_o}\right)r^*$ we end up with
\begin{eqnarray}
\label{2.39}
ds^2 = & &\left(1+{16\pi \eta^2\over \phi_o(2\omega +3)}\ln{r\over r_o}\right)\biggr[dt^2-dr^2-\left(1-{8\pi \eta^2\over \phi_o}\right)\nonumber \\
& & \times r^2(d\theta^2 + \sin^2\theta d\varphi^2)\biggr].
\end{eqnarray}

In order to obtain the correct Newtonian limit from Brans-Dicke field equations the constant $\phi_o$ must be given by \cite{7} $\phi_o=\left({2\omega +4\over 2\omega +3}\right){1\over G}$. Then, the final form of (\ref{2.39}) reads 

\begin{eqnarray}
\label{2.40}
ds^2 = & &\left(1+{16\pi \eta^2G\over (2\omega +4)}\ln{r\over r_o}\right)\biggr[dt^2 -dr^2-\biggr(1- 8\pi \eta^2G\left({2\omega +3\over 2\omega +4}\right)\biggr)
\nonumber \\
& &\times r^2(d\theta^2 + \sin^2\theta d\varphi^2)\biggr].
\end{eqnarray}

Thus, we have shown that in the weak field approximation equation (\ref{2.40}) represents the space-time generated by a global monopole in Brans-Dicke theory of gravity. Analogously to the General Relativity case this curved space-time also presents a deficit solid angle in the hypersurfaces $t=const.$ The area of a sphere of radius $r$ in these spaces would be given by 
$$
4\pi r^2\left[1- 8\pi \eta^2G\left({2\omega +3\over 2\omega +4}\right) + {16\pi \eta^2 G\over (2\omega +4)}\ln{r\over r_o}\right] \hspace{0.2cm},
$$ 
rather than $4\pi r^2$.

Also, a simple comparison of (\ref{2.40}) with Barriola-Vilenkin solution shows that for large values of $\omega$ both space-times are related by a conformal transformation. In this case the motion of light rays is the same in the two space-times. For finite values of $\omega$, null geodesics in the space-time of Brans-Dicke global monopole are still closely related to their counterpart in General Relativity. Indeed, the only change predicted by Brans-Dicke theory reduces, in this case, to the replacement of the Newtonian gravitational constant $G$ by the $\omega$-dependent ``effective'' gravitational constant 
$G_{0}=\frac{2\omega +3}{2\omega+4}G$. For a value of $\omega$ consistent with solar system observations, say, $\omega\sim 500$ \cite{8}, it would mean that massless particles travelling in the space-time described by (\ref{2.40}) would experience a gravitational strength $G_{0}\sim 0,999 G$.

\renewcommand{\thesection}{\Roman{section}.}

In conclusion we see that in going from General Relativity to Brans-Dicke theory both space-time curvature and topology are affected by the presence of the scalar field. In particular the deficit solid angle becomes $\omega$ dependent. As a consequence, following Barriola-Vilenkin's argument concerning light propagation in the gravitational field of a global monopole one can easily show that a light signal propagating from a source $S$ to an observer $O$ when $S,O$ and the monopole are perfectly aligned produces an image with the form of a ring of angular diameter given by 
\begin{eqnarray}
\delta \Omega = 8\pi^2 \eta^2 \left({2\omega + 3\over 2\omega +4}\right)G{l\over l +d}\hspace{.5cm}, \nonumber
\end{eqnarray}
where $d$ and $l$ are the distances from the monopole to the observer and to the source, respectively.

Another interesting physical property in connection with Brans-Dicke's global monopole involves the appearance of gravitational forces  exerted by the monopole on the matter around it. This effect is absent in the case of General Relativity's monopole as was shown in ref. \cite{3}. To see how this gravitational effect comes about one has to work out the Newtonian potential associated with (\ref{2.40}). As is well known in Galilean coordinates the motion of a nonrelativistic test particle in a weak gravitational field is given by the equation \cite{9}

\begin{eqnarray}
\label{2.41}
\mbox{\"x}^{i}=-\frac{1}{2}\frac{\partial h_{00}}{\partial x^{i}},
\end{eqnarray}
where $g_{\mu\nu}=\eta_{\mu\nu}+h_{\mu\nu}$ and $\eta_{\mu\nu}=\mbox{diag}(1,-1,-1,-1)$ is Minkowski metric tensor. In order to express (\ref{2.40}) in Galilean coordinates let us consider the transformation 

\begin{eqnarray}
t=\left[1-4\pi\eta^{2}G\left(\frac{2\omega+3}{2\omega +4}\right)\right]T,
\nonumber
\end{eqnarray}

\begin{eqnarray}
r=\left[1+4\pi\eta^{2}G\left(\frac{2\omega+3}{2\omega +4}\right)-
4\pi\eta^{2}G\left(\frac{2\omega+3}{2\omega +4}\right)\ln{\frac{R}{R_{0}}}\right]R,
\nonumber
\end{eqnarray}
with

\begin{eqnarray}
R_{0}=\left[1-4\pi\eta^{2}G\left(\frac{2\omega+3}{2\omega +4}\right)\right]r_{0}.
\nonumber
\end{eqnarray}
Then, we have

\begin{eqnarray}
\label{2.42}
ds^{2}&=&\left[1-8\pi\eta^{2}G\left(\frac{2\omega+3}{2\omega +4}\right)+
\frac{16\pi\eta^{2}G}{2\omega+4}\ln{\frac{R}{R_{0}}}\right]dT^{2}
\nonumber \\
& &-\left[1-8\pi\eta^{2}G\left(\frac{2\omega+1}{2\omega +4}\right)\ln{\frac{R}{R_{0}}}\right]
(dx^{2}+dy^{2}+dz^{2}),
\end{eqnarray}
with $R=[x^{2}+y^{2}+z^{2}]^{1/2}$. Thus, (\ref{2.41}) becomes, finally

\begin{eqnarray}
\label{2.43}
\mbox{\"x}^{i}=-\frac{4\pi\eta^{2}G}{(\omega+2)}\frac{x^{i}}{R^{2}},
\end{eqnarray}
which shows explicitly that particles around the monopole are subject to an attractive force exerted by it.

Naturally, if it turns out to be that global monopoles possess any kind of physical reality then a number of other effects such as quantum particle creation \cite{10}, vacuum polarization \cite{11} and gravitational scattering \cite{12} among others, which would be in principle amenable to observation may be investigated with the help of equation (\ref{2.40}), thereby providing alternative ways for testing the predictable power of both General Relativity and Brans-Dicke theory. 

C. Romero was partially supported by CNPq (Brazil).

\end{document}